\documentclass[prd,aps,superscriptaddress,twocolumn,floatfix,10pt]{revtex4-2}
\usepackage{amssymb,amsmath,amsthm,latexsym,mathrsfs,verbatim,
            mathtools,needspace,enumitem,etoolbox,graphicx,physics,
            microtype,afterpage,xspace,tabularx,lmodern,multirow,slashed}
\usepackage{graphicx}
\usepackage{environ}
\usepackage{csvsimple,etoolbox}
\usepackage{tikz}
\usepackage{tensor}
\usepackage{placeins}
\usepackage{xspace}
\usepackage{orcidlink}
\usepackage{stmaryrd}
\usepackage{xcolor}
\usepackage{float}
\usepackage{textcomp, gensymb}
\usepackage{flushend} 
\usepackage{comment}
\usepackage{acronym}
\newacro{CIVP}{Characteristic Initial Value Problem}
\newacro{CFL}{Courant-Friedrichs-Levy}
\newacro{CFL}{Self-Similarity Horizon}
\newacro{DSS}{Discretely Self-Similar}
\newacro{EFEs}{Einstein Field Equations}
\newacro{SSH}{Self-Similarity Horizon}
\newacro{GR}{General Relativity}
\newacro{IST}{Instituto Superior Técnico}

\newcommand{\rita}[1]{{\textcolor{orange}{\sf{[Rita: #1]}}}}

\usepackage{hyperref}
\usepackage[hang,small,bf,tight]{subfigure}
\usepackage[normalem]{ulem}
\usepackage{adjustbox}

\begin{document}

\title{Characteristic Critical Collapse of a Yang-Mills Field With Null Infinity}
\author{Rita P. Santos\orcidlink{0009-0002-2952-7431}} 
\affiliation{CENTRA, Departamento de F\'{\i}sica, Instituto Superior T\'ecnico -- IST, Universidade de Lisboa -- UL, Avenida Rovisco Pais 1, 1049-001 Lisboa, Portugal}
\affiliation{Institut de Ci\`encies de l'Espai (ICE, CSIC), Campus UAB, Carrer de Can Magrans s/n, 08193 Cerdanyola del Vall\`es, Spain}
\author{Krinio Marouda\orcidlink{0000-0003-1030-8853}} 
\affiliation{CENTRA, Departamento de F\'{\i}sica, Instituto Superior T\'ecnico -- IST, Universidade de Lisboa -- UL, Avenida Rovisco Pais 1, 1049-001 Lisboa, Portugal}
\author{David Hilditch\orcidlink{0000-0001-9960-5293}} 
\affiliation{CENTRA, Departamento de F\'{\i}sica, Instituto Superior T\'ecnico -- IST, Universidade de Lisboa -- UL, Avenida Rovisco Pais 1, 1049-001 Lisboa, Portugal}
\date{\today}

\begin{abstract}
Solutions to the Einstein equations near the threshold of black hole formation exhibit remarkable behavior known as critical phenomena gravitational collapse. In this work we perform characteristic evolution in compactified Bondi coordinates in order to study the critical collapse of a Yang-Mills field, allowing for the extraction of global quantities such as the Bondi mass and news function. Our numerical approach is fourth-order accurate. First, we demonstrate that the collapsing field exhibits local \ac{DSS} behavior, characterized by an echoing period of~$\Delta \simeq 0.7388$, agreeing with previous works up to the second decimal place. We find that global quantities such as the Bondi mass and news function display the same \ac{DSS} behavior. We furthermore show that the mass of the black holes formed during near-threshold evolutions scales as a function of the distance to the critical parameter, with a critical exponent of approximately~$\gamma=0.1977\pm0.0009$. Finally, our findings indicate that these results are universal, irrespective of the initial data.
\end{abstract}

\maketitle

\section{Introduction}
\label{intro}

Critical phenomena in gravitational collapse were first discovered by Choptuik~\cite{choptuik1993}, who studied the dynamical evolution of one-parameter families of initial data that describe a real massless scalar field in spherical symmetry. It was found that the evolution of such data asymptote generically to one of two possible end-states. If initial data is strong enough, a black hole will form, whereas sufficiently weak data will disperse to leave behind flat spacetime. Close to the threshold of collapse however, Choptuik found a very interesting behavior. Despite the complicated form of the \ac{EFEs}, at the threshold of black hole formation the dynamics of the system become relatively simple and exhibit universal behavior in which, as described below, the details of the specific initial data are insignificant. In the context of spherical symmetry, similar behavior has been observed with various different matter models~\cite{Gundlach_2007, Gundlach:2025yje}. Universal solutions that lie at the threshold are called the critical solution. 

The picture that emerges from this work is that, at least for many matter models in the spherical context, near the threshold between black hole formation and dispersion, solutions to the \ac{EFEs} show features of the critical solution, such as power-law scaling with respect to the distance to the threshold, self-similarity and universality. Firstly, scaling refers to the fact that the global maximum of quantities~$l$ with units of length, in barely subcritical runs, depend on a parameter~$p$ of a family of initial data and that this relation follows a power-law
\begin{align}
    l \simeq |p-p_*|^{\gamma}\,,
    \label{eq:massscaling}
\end{align}
where the critical value~$p_*$ depends on the particular one-parameter family of initial data. Similar behavior, also referred to as scaling, can be observed in certain quantities with the same units on the barely supercritical runs. The critical exponent~$\gamma$ is often found to be universal for a given matter model. Secondly, near-critical evolutions are approximately self-similar for a while before dispersion or collapse. In the particular case of \ac{DSS}, a case which is realized in the massless scalar field model, up to a conformal rescaling the solutions are found to be periodic in similarity time with a period~$\Delta$. Finally, it is found that near-critical solutions look the same for some time of the evolution. This is what is referred to as universality. For example, in the sub-critical side, the more we tune our initial data to the critical parameter, the longer this evolution will show the features of the critical solution, before dispersing to flat spacetime. This means besides the distance from the threshold of black hole formation, all details of the initial data are irrelevant. This means that, as well as power-law values~$\gamma$, parameters such as the period~$\Delta$ are universal.

In the case that the super-critical evolutions near the threshold lead to the formation of infinitesimal black holes, the critical collapse is characterized as type-II, which is the type we are concerned with in this work. On the other hand, in type-I critical collapse there is a discontinuity in the mass scaling, and it is not possible to form black holes with mass smaller than a certain universal value dependent on physical scale set by the model.

In this work, we restrict to spherical symmetry and study solutions to the Yang-Mills matter model minimally coupled to \ac{GR} near the threshold of collapse. It is known that, depending on the initial data and field ansatz considered, the Yang-Mills field collapse can display types I or II behavior~\cite{Choptuik_1996, Maliborski_2018}. The general spherically symmetric Yang-Mills connection has two free potentials, but most numerical work so far has considered the so-called magnetic ansatz, in which one of potentials is set to zero. Here, we focus on the type-II collapse withing the purely magnetic ansatz. Our aim is to explore critical collapse using compactified null coordinates. While the echoing period in the \ac{DSS} critical solution for the scalar field is~$\Delta \simeq 3.44$, in the Yang-Mills field case, we have~$\Delta \simeq 0.7$~\cite{Gundlach_1997}. At a practical level this means that the Yang-Mills field allows for a less computationally demanding study of critical collapse.

The code developed is based on the solution of a \ac{CIVP}, in which the initial data is specified on a null hypersurface. Our spacetime is then foliated by a family of non-intersecting outgoing null hypersurfaces. We compactify our domain in order to include both the origin and future null-infinity~$\mathscr{I^+}$ in our computational domain. Our motivation for this choice of evolution is that it allows us to study the collapse from the point of view of~$\mathscr{I^+}$, corresponding to an idealized astrophysical observer. In this manner, our choice of numerical setup allows us to study global quantities such as the Bondi mass and the news function familiar from discussions of gravitational waves in asymptotically flat spacetimes.

Different aspects of the gravitational dynamics of a Yang-Mills field have been studied with different continuum formations and numerical approximations. For instance, the system has been studied as a \ac{CIVP}, with an evolution based on compactified null hypersurfaces~\cite{Purrer_2009}, with a focus on the late-time behavior of subcritical evolutions far from the critical regime. Again focusing on late-time tails, results have been presented using compactified hyperboloidal evolution~\cite{Rinne_20102}. In their setup the foliation is spacelike but leaves nevertheless terminate at~$\mathscr{I^+}$. Critical collapse with the model was first studied in~\cite{Choptuik_1996, Choptuik_1999} using a Cauchy formulation. Building on~\cite{Rinne_20102}, critical collapse of a Yang-Mills field has been examined in~\cite{Rinne_2014} with a hyperboloidal foliation, with a particular focus on type-III collapse, a rare finding in which the two final-states on both sides of the threshold are black holes, but the Yang-Mills field is in different vacuum states. The state-of-the-art was provided in Schwarzschild-like coordinates by~\cite{Maliborski_2018}. For the type-II collapse of the magnetic ansatz, they find a critical exponent of~$\gamma=0.19714\pm0.00074$ and an echoing period of~$\Delta=0.7364\pm0.0007$ which, we will see, are compatible with our findings. 

In~\cite{Gundlach_2019}, the threshold of collapse for two interacting spherical fields, a massless scalar field and a Yang-Mills field, was studied. Based on numerical results the existence of a “quasi-discretely self-similar” solution shared by the two fields was conjectured. Empirically this solution is equal to the Choptuik solution at infinitely small scales and the type-II Yang-Mills critical solution at large scales, with a gradual transition from one to the other. This work employs the use of single-null coordinates, but in an non-compactified domain, not allowing for radiation studies at $\mathscr{I^+}$.

Using a complementary strategy to the hyperboloidal approach, global aspects of critical collapse of a scalar field were studied in~\cite{Purrer_2005}. A self-gravitating massless scalar field in spherical symmetry was evolved numerically with a characteristic formulation, using a compactified grid and thus including $\mathscr{I^+}$ in the computational domain.

In view of the above, in this work we follow the formulation of~\cite{Purrer_2005} but make adjustments to achieve fourth-order accuracy on compactified null-slices, which allows efficient computation of radiative quantities at~$\mathscr{I^+}$. The novelty consists of the combination of a characteristic formulation and the study of the radiation emitted during a Yang-Mills field collapse. The advantage is not only that it allows for the study of~$\mathscr{I^+}$, but also the fact that the evolution equations, at least in the spherical setting, are much simpler than in other formulations such as hyperboloidal or Cauchy. The study of solutions close to criticality then becomes very computationally efficient, allowing for a more straightforward tuning to the critical solution. A characteristic approach also simplifies the study of the causal structure of our solution and, as observed in~\cite{Garfinkle:1994jb} can be used to side-step the need for mesh refinement. Recent further discussion of characteristic formulations can be found in~\cite{Gundlach:2024mld}.

The paper is organized as follows. In section~\ref{continuummodel} we present our formulation the field equations. In section~\ref{numericalmethod} we presents our numerical approach. Our results concerning critical collapse are presented in section~\ref{sec:numericalresults}. Finally, we conclude in section~\ref{sec:conclusions}. Geometric units are used throughout.

\section{Continuum Model}
\label{continuummodel}

In this study we focus on the threshold of collapse of a purely magnetic SU(2) Yang-Mills field in spherical symmetry. We begin in this section by formulating the equations of motion for this model along the lines of~\cite{Rinne_2014, Gundlach_2019}.

Our work differs from these as we consider a characteristic foliation of spacetime with compactified Bondi coordinates, as opposed to the non-compactified form used in~\cite{Gundlach_2019} or the hyperboloidal evolution of~\cite{Rinne_2010, Rinne_2014}. Ultimately this has the advantage that we do not need to solve constraints for initial data but can resolve the dynamics out to infinity. Our continuum model follows the characteristic formulation of \ac{GR} previously used in~\cite{Purrer_2005} to study the collapse of a massless scalar field. Their work uses compactified Bondi coordinates, which simplifies the study of the region close to the center of spherical symmetry. Their approach allows for the study of the gravitational collapse from the point of view of an observer at future null-infinity, which we now utilize to study the collapse of a purely magnetic SU(2) Yang-Mills field. Since the period of the critical solution is significantly smaller in comparison with that of the massless scalar field, we are able to calculate nearby solutions accurately without the use of either mesh-refinement or infalling coordinates with regridding. Our adjustments to the numerical approach of~\cite{Purrer_2005} are discussed below in section~\ref{numericalmethod}.

\subsection{Geometry}

The line element in Bondi coordinates $\{u,r,\theta,\phi\}$ is of the form
\begin{align}\label{eq:line-element}
\begin{aligned}
    \textrm{d}s^2 = & -e^{2 \beta(u,r)}\frac{V(u,r)}{r}\textrm{d}u^2 
    -2e^{2 \beta(u,r)}\textrm{d}u\textrm{d}r \\
     & + r^2(\textrm{d}\theta^2+\sin^2\theta \textrm{d}\phi^2),
\end{aligned}
\end{align}
in which~$\beta(u,r)$ and~$\frac{V(u,r)}{r}$ are smooth metric functions. Our gauge choice will be such that our outgoing null slices are parameterized by the proper time of an observer located at the origin, where the strong field dynamics are happening.

In order to obtain a fully regular system of evolution equations, we work with the Misner-Sharp mass function~\eqref{eq:mass}. In this way, we eliminate~$V$ by using
\begin{align}\label{eq:mass}
    m(u, r)=\frac{r}{2}\left[1-\frac{V}{r} e^{-2 \beta}\right].
\end{align}
In numerical studies of \ac{GR}, it is often useful to perform a compactification of the radial coordinate~$r$. We perform a coordinate transformation which maps a half-line~$[0,\infty)$ to a finite segment~$[0,1]$, according to
\begin{align}\label{eq:compacc}
    x:=\frac{r}{1+r}.
\end{align} 
Points at~$\mathscr{I^+}$ are then included in our grid at~$x=1$. In such a manner, we can simulate observers at null-infinity which will allow us to extract global properties of our problem. 

\subsection{Yang-Mills Field Collapse}
\label{ymequationss}

As in~\cite{Gundlach_2019}, the purely magnetic Yang-Mills field in spherical symmetry can be parametrized as
\begin{align}
    F=\textrm{d}W \land (\tau_1 \textrm{d}\theta + \tau_2 \sin \theta \textrm{d} \phi) 
    -(1-W^2)\tau_3 \textrm{d}\theta \land \sin\theta \textrm{d}\phi,
\end{align}
where~$\tau_i$ are the Pauli matrices with $\textrm{tr}(\tau_i\tau_j) = \delta_{ij}$. (See also \cite{Choptuik_1996}, which used a similar parametrization to study the critical collapse of a Yang-Mills field in Cauchy coordinates.) In the presence of a Yang-Mills field~$W$ the Einstein Equations are
\begin{align}\label{eq:eisnteineqym}
    G_{ab}=8\pi T^{(W)}_{ab},
\end{align}
with
\begin{align}
\begin{aligned}
    T^{(W)}_{ab} &= \Tilde{T}^{(W)}_{ab} - \frac{1}{4} g_{ab} \Tilde{T}^{(W)}, \\
    \Tilde{T}^{(W)}_{ab} & = \text{diag}(2r^{-2}\nabla_a W \nabla_b W,Pr^2\gamma_{ab}),\\
    P &= r^{-2} \nabla_a W \nabla^a W + r^{-4}(1-W^2)^2,
\end{aligned}
\end{align}
where~$\gamma_{ab} = \textrm{diag}(1, \sin^2\theta)$ is the unit metric on the 2-sphere, and~$P$ has units of pressure. Observe that~$\Tilde{T}^{(W)}_{ab}$ is block diagonal in the~$uu$, $ur$ and~$rr$ components, and diagonal in the rest. The Einstein Equations result in the hypersurface equations
\begin{align}\label{eq:both-ym}
\begin{aligned}
    \beta_{,r} &= 4 \pi \frac{W_{,r}^2}{r} \\
    m_{,r} &= \frac{2 \pi}{r^2}((W^2-1)^2+2 r (r-2 m) W_ {,r}^2).
\end{aligned}
\end{align}

Following \cite{Gundlach_2019}, $W$ obeys the wave equation in the 2-dimensional \textit{ur} plane (the angular dimensions are suppressed due to spherical symmetry), with a potential term, namely
\begin{align}\label{eq:wave-YM}
\begin{aligned}
    \square_h W =&  e^{-2 \beta}\left[ \left(\frac{V}{r}\right)_{, r} \partial_r-2 \partial_u \partial_r+\frac{V}{r} \partial_{r r}\right] W \\
    =& -\frac{W(1-W^2)}{r^2},
\end{aligned}
\end{align}
in which~$\square_h$ denotes the two dimensional D'Alembert operator, which is given by
\begin{align}
\square_h = e^{-2 \beta}\left[ \left(\frac{V}{r}\right)_{, r} \partial_r-2 \partial_u \partial_r+\frac{V}{r} \partial_{r r}\right].
\end{align}
Using~\eqref{eq:mass}, Equation~\eqref{eq:wave-YM} can be rewritten as
\begin{align}\label{eq:YMevol4pre}
    W_{,ur} &= \frac{e^{2 \beta}}{2 r^2}\big(W-W^3+2 W_{,r}
    (m-r m_{,r}+r (r-2 m) \beta_{,r})\nonumber\\
    &\qquad\quad +r (r-2 m) W_{,rr}\big).
\end{align}
We can then use Equation~\eqref{eq:YMevol4pre} to perform our time evolution and find our field~$W$ on slices of constant~$u$. In fact, we introduce a parametrization for~$W$ such that we regularize Equation~\eqref{eq:YMevol4pre}. An appropriate ansatz for~$W$~\cite{Choptuik_1996,Gundlach_2019} is given by
\begin{align}\label{eq:w-to-chi}
    W=1+\left(\frac{r}{1+r}\right)^2 \chi \equiv 1+ x^2 \chi,
\end{align}
in which~$\chi(u,x) = \mathcal{O}(1)$ both at the origin and at~$\mathscr{I^+}$. To make the numerical implementation more straightforward, we define a new variable~$\xi(u,x)\equiv x^2 \chi (u,x)$. 

Our final evolution system of equations is then composed of two Einstein equations~\eqref{eq:both-ym} and the wave equation originating from~$W$~\eqref{eq:YMevol4pre}, regularized by the parametrization given by~\eqref{eq:w-to-chi}.

\subsection{Asymptotic Quantities}
\label{asymptotic-quantities}

We now define the radiation quantities at~$\mathscr{I^+}$ for the collapse, namely the Bondi mass and the news function.

\subsubsection{Bondi Mass}

If we define the total energy enclosed by a surface, this energy definition is said to be quasilocal. The Misner-Sharp mass, given by Equation~\eqref{eq:mass}, is one such quantity. On the other hand, the Bondi mass is a global mass definition for asymptotically flat spacetimes. It captures the mass that remains on a null hypersurface of constant~$u$, and is defined as
\begin{align}
    M(u)=\lim_{r\to\infty} m(u,r)|_{u=\textrm{const}}.
\end{align}
Because this hypersurface doesn't intersect any of the radiation emitted prior to the retarded time~$u$, it turns out that the Bondi mass can only decrease with increasing retarded time~$u$. For instance, in an isolated system, outgoing waves can radiate physical energy to~$\mathscr{I^+}$. The Einstein equation in~$\{rr\}$ gives us an expression for the mass change that can be written as

\begin{align}\label{eq:massloss}
\begin{gathered}
  m_{,u} = -4 e^{-2 \beta} \pi r^2 \; T_{uu}(u,r) \\ + 8 \pi r \; (r-2 m) \; T_{ur}(u,r) \\
  - 2 \pi e^{2 \beta} (r-2m)^2 \; T_{rr}(u,r).
\end{gathered}
\end{align}
Equation~\eqref{eq:massloss} is valid for a general matter model in spherical symmetry. For the case of a Yang-Mills field, the mass loss formula at $\mathscr{I^+}$ is simply
\begin{align}\label{eq:masslossanalyticalym}
    m_{,u}=-8 \pi e^{-2 \beta} \xi_{,u}^2,
\end{align}
in which~$u$ is equal to the central time~$u_C$. We can conclude that the Bondi mass is in general not conserved, being monotonically decreasing in~$u$. This is as expected physically since no mass can enter our compactified domain.

\subsubsection{News Function}

The news function~$N(u)$ is a quantity related to the emission of energy momentum through $\mathscr{I^+}$. The Bondi mass-loss equation provides a relation between~$N(u)$ and the Bondi mass~$M(u)$~\cite{Purrer_2009}. It can be written as
\begin{align}\label{eq:newsss}
    e^{-2H(u_C)}\frac{\textrm{d}M(u_C)}{\textrm{d}u_C} = -4 \pi N^2(u_C),
\end{align}
in which~$H=\beta(u_C,\infty)$. As discussed in~\cite{Purrer_2007}, $H$~is related to the redshift. It can be argued that if~$H\rightarrow\infty$, then a finite amount of central time~$u_C$ corresponds to an infinite amount of time~$u_B$ (see Equation~\eqref{eq:time_relation}). This means that light rays being emitted from the center are infinitely redshifted.

We denote the Bondi time by~$u_B$. In~\cite{Purrer_2007}, it is shown that the relation between central and Bondi time in the limit~$r\rightarrow\infty$ is
\begin{align}\label{eq:time_relation}
    \textrm{d}u_B = e^{2H}\textrm{d}u_C.
\end{align}
Using the mass loss formula we derived for the Yang-Mills system~\eqref{eq:masslossanalyticalym} and the Bondi mass-loss Equation~\eqref{eq:newsss}, we find an expression for the News at~$\mathscr{I^+}$ for the case of this physical system
\begin{align}
    N(u_C) = \sqrt{2} \xi_{,u_C}.
\end{align}
This can be translated onto Bondi time as
\begin{align}
    N(u_B) = \sqrt{2} e^{2\beta} \xi_{,u_B}.
\end{align}

\section{Numerical Method}
\label{numericalmethod}

As in~\cite{Purrer_2005} we solve the \ac{CIVP} numerically, meaning that we use a null foliation of spacetime and specify our initial data on a hypersurface of constant~$u$. The main difference here, besides the fact that they study a scalar field critical collapse, is that we use a different integration scheme. Pürrer et al.~\cite{Purrer_2005} integrate over a null parallelogram, developing a code based on a Diamond Integral Characteristic Evolution~\cite{Husa_2000}. We instead use a Runge-Kutta integrator for both the time~$u$ and space~$x$ integrations, which are done separately. Our integration scheme is straightforward to apply and has the advantage of being globally fourth-order accurate. We started by implementing a second-order accurate method and found, somewhat unsurprisingly, that our results improved substantially once we upgraded to fourth-order accuracy. Derivatives were approximated using a fourth-order finite difference scheme given by
\begin{align}\label{eq:finite-diff}
    u'=\frac{u_{i-2}-8u_{i-1}+8u_{i+1}-u_{i+2}}{12h} + \frac{h^4}{30}u^{(5)}.
\end{align}
At the first and second gridpoints of our grid, our finite difference operator is not defined, thus we need to use a different stencil, lopsided by two and one gridpoints respectively, and we
need to make sure that the error terms match. Following~\cite{alcubierre_2006}, we use lopsided derivative operators, which are found to be
\begin{align}\label{eq:finite-diff-matched}
\begin{aligned}
u'&\approx\frac{-2u_{i-1}-15u_{i}+28u_{i+1}-16u_{i+2}+6u_{i+3}-u_{i+4}}{12h},\\
u'&\approx\frac{-27u_{i}+58u_{i+1}-56u_{i+2}+36u_{i+3}-13u_{i+4}+2u_{i+5}}{12h},\\
\end{aligned}
\end{align}
which have the same leading error as Equation~\eqref{eq:finite-diff}. We use Kreiss-Oliger dissipation to damp the formation of high frequency modes.

The code developed for this project (a combination of \texttt{Julia}, \texttt{Python} and \texttt{Mathematica} files) can be found at the \href{https://github.com/rita-png/yangmills-criticalcollapse}{GitHub repository}.

\section{Numerical Results}
\label{sec:numericalresults}

We implemented the evolution system of equations~\eqref{eq:both-ym} and~\eqref{eq:YMevol4pre} as derived in Section~\ref{ymequationss}. In this section we begin with a description of given data and diagnostics, before briefly presenting a representative validation test and finally moving on to discuss our physical results.

\subsection{Initial Data and Diagnostics}

We construct initial data families for~$\xi$ that are Gaussian-like:
\begin{align}\label{eq:initial-dataym}
    \xi(u_0,x) = \mathrm{A} r(x) \exp\left[-\left(\frac{r(x)-r_0}{\sigma}\right)^2\right].
\end{align}
We fix~$r_0=0.3$ and~$\sigma=0.1$, and study the influence of changing the amplitude~$A$, our tuning parameter.

The critical solution is parameterized by a critical amplitude~$\mathrm{A}=\mathrm{A_*}$ which determines the separation between two different scenarios in parameter space: supercritical solutions, which correspond to black hole formation, and subcritical evolutions, in which the initial data eventually disperses to flat spacetime. We find the critical amplitude that produces such evolutions by performing a bisection search in the parameter~$\mathrm{A}$. From each bisection, we extract relevant quantities of critical solutions, such as the self-similarity echoing period~$\Delta$ and the accumulation time~$u_*$. For a near-critical solution, the approximate value of the accumulation time defines an approximate location (advanced time) of the \ac{SSH}.

In practice, we detect the formation of an apparent horizon by monitoring the value of the compactness~$2m/r$ throughout the evolution, with $2m/r=1$ signifying the presence of an apparent horizon. In our particular null coordinates, our slices do not penetrate apparent horizons. Therefore, we \textit{soften} the condition by using~$2m/r>0.71$ as the condition to mark evolutions as supercritical, as is common practice. We have adjusted the specific value chosen for this criteria and find that within some range this has little effect on the outcome.

Evolutions closer to criticality are required to `slow down' according to a \ac{CFL}-type condition, which is a condition that guarantees that the full numerical domain of dependence contains the physical domain of dependence. This condition can be written as
\begin{equation}
    \mathrm{C}=\frac{v \Delta t}{\Delta x}\leq 1,
\end{equation}
in which~$\mathrm{C}$ is the Courant number and~$v=\frac{dx}{du}$. The value of~$v$ can be obtained from Equation~\eqref{eq:line-element} by setting $\textrm{d}\theta=\textrm{d}\phi=0$, and can be rewritten in terms of~$m$ and~$x$ as
\begin{equation}
    \frac{\textrm{d}x}{\textrm{d}u}=\frac{(1-x)^3 e^{2\beta}(2m/x-\frac{1}{1-x})}{2}.
\end{equation}
We start the evolution with~$\Delta u = 0.5 \; \Delta x$, updating~$\Delta u$ at each iteration according to~$\Delta u = 0.5 \; \Delta x \left(\frac{\textrm{d}x}{\textrm{d}u}\right)^{-1}$.
This ensures that the distance any information travels during a timestep~$\Delta t$ must be smaller than the distance between mesh elements~($\Delta x$). Our scheme is not strictly causal but this poses no difficulties beyond this (expected) restriction on the timestep.

As we tune closer to~$\mathrm{A}_*$ our solutions are expected to exhibit critical behavior. One of the characteristics is discrete self-similarity of the collapsing field solution. This self-similarity is characterized by an echoing period~$\Delta$, which we will extract from our evolutions.

Based on our Bondi coordinates $\{u,r,\theta,\phi\}$, we define the specific \ac{DSS}-adapted time coordinate as
\begin{align}
    T=-\ln(u_*-u), %
\end{align}
where~$u$ is the proper time at the origin, and~$u_*$ is the the accumulation time, at which the curvature diverges in the limit of infinite tuning to the threshold. $T$ is then what we will call the similarity time, which is defined for a constant~$u_*>0$ and~$u<u_*$ \cite{Baumgarte_2018}.

The accumulation time $u_*$ can be estimated by taking two pairs of consecutive zero-crossings of the magnitude of the collapsing field at the origin, ($u_n$,$u_{n+1}$) and ($u_m$,$u_{m+1}$). Assuming that each pair differs in half of the period $\Delta/4$, we can solve for the accumulation time \cite{Baumgarte_2018} obtaining
\begin{align}
    u_* = \frac{u_n u_{m+1}-u_{n+1}u_m}{u_n-u_{n+1}-u_m+u_{m+1}}.
\end{align}

The same reasoning can be applied to provide an estimate for the echoing period $\Delta$ \cite{Baumgarte_2018} resulting in
\begin{align}\label{eq:deltaapprox}
    \Delta=2 \ln\left(\frac{u_*-u_n}{u_*-u_{n+1}}\right).
\end{align}
Each self-similar repetition of the solution at progressively smaller scales is referred to as an echo.

\begin{figure}[t!]
\centering
\includegraphics[width=\columnwidth]{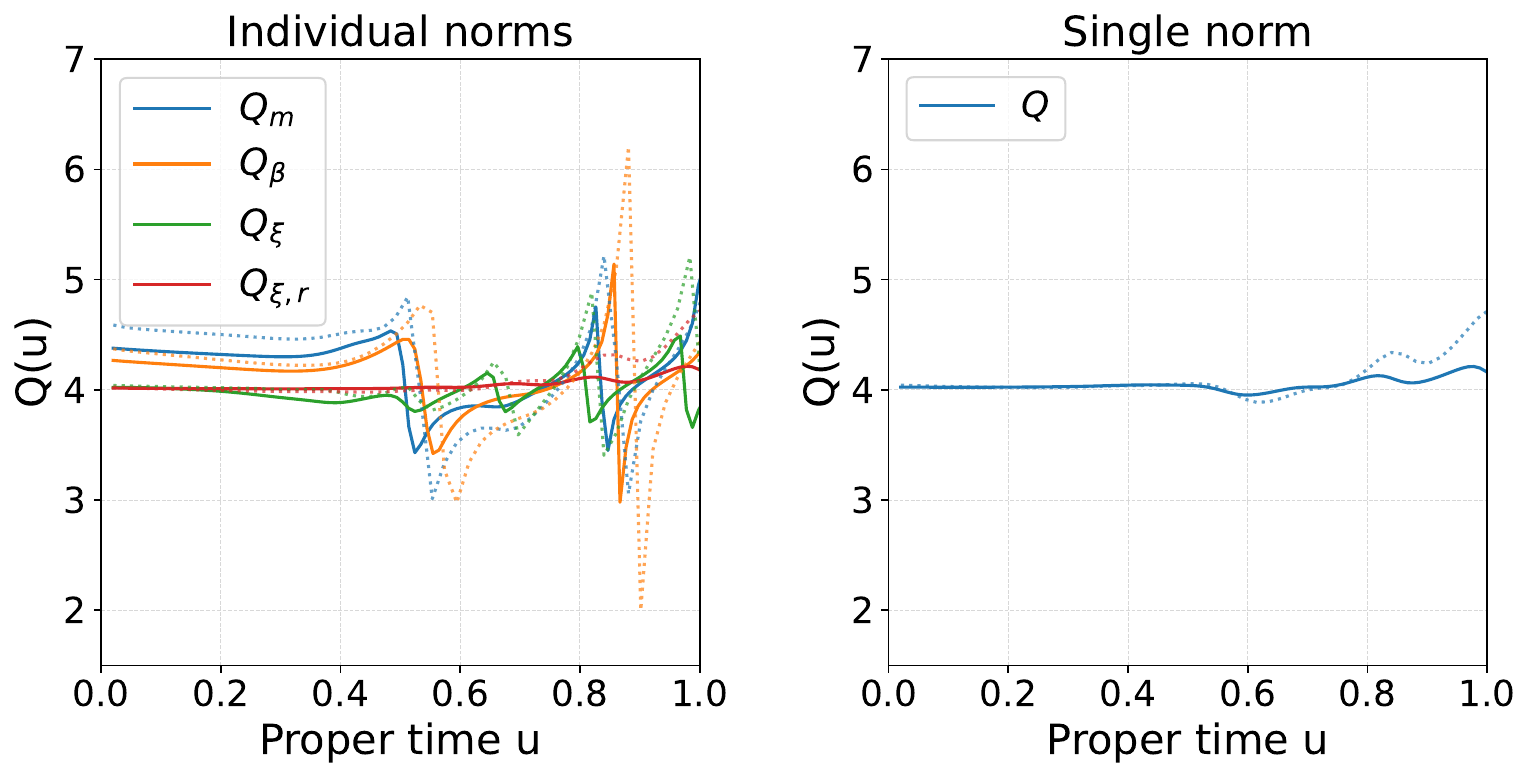}
\caption{Convergence factor~$Q(u)$ throughout the evolution, using a code that is fourth-order accurate. The dotted and solid lines correspond to a grid with~$100$ and~$200$ points at the lower resolution, respectively. We see that as we increase resolution, $Q$ gets closer to 2. The initial data is constructed using Equation~\eqref{eq:initial-dataym}, and setting~$A=0.01$, $r_0=0.3$ and~$\sigma=0.08$. We stop the convergence test when the magnitude of all fields is smaller than~$10^{-7}$. We use a lower resolution here than in our near-critical runs, but as we increase resolution and data strength, the solution converges in a similar manner.\label{fig:convergence}}
\end{figure}

\subsection{Convergence Tests}

To validate our numerical solutions we perform convergence tests. We compute the solution~$f$ at three different resolutions~$\Delta_1$, $\Delta_2$, and~$\Delta_3$, with~$\Delta_3=\Delta_2/2=\Delta_1/4$. The convergence factor~$\mathrm{Q}$ is defined as~$\mathrm{Q(u)}=\frac{||f_{\Delta_1}-f_{\Delta_2}||}{||f_{\Delta_2}-f_{\Delta_3}||}$, with~$||\cdot||$ a discrete approximation to the~$L^2$ norm, which we evaluate throughout the evolution for each variable separately, as shown in the left panel of Figure~\ref{fig:convergence}. Since all the numerical methods we implement are fourth-order accurate, we expect~$\mathrm{Q}$ to remain close to~$\sim4$ throughout the evolution. Global zero-crossings of a particular variable cause jumps in the apparent convergence rates, and so a more informative plot is given by the overall code convergence rate, obtained by computing the norm taking into account all the variables together, which is shown on the right panel of the same figure. We stop the convergence test once the variables disperse and get to values close to the numerical precision, which happens at~$u=1.0$. At this point, we can no longer trust the computed convergence factor. In Figure~\ref{fig:convergence}, we see that as we increase resolution the convergence factor~$Q$ gets closer to$\sim4$, indicating good fourth-order convergence as desired.

\subsection{Identification of Critical Behavior}
\label{criticym}

\begin{figure}[t!]
	\centering
	\subfigure{\includegraphics[width=0.75\columnwidth]{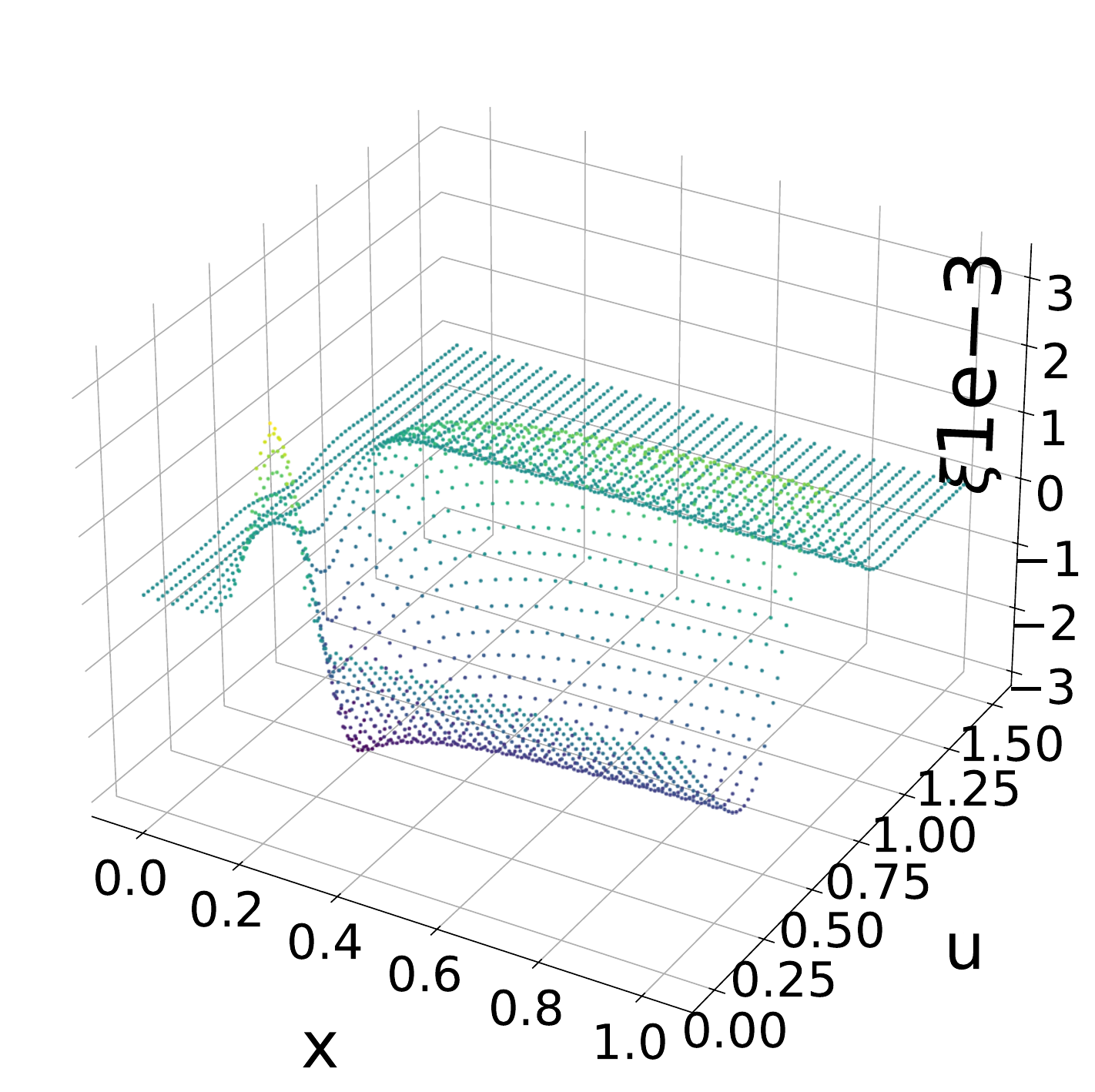}}\qquad
	\subfigure{\includegraphics[width=0.75\columnwidth]{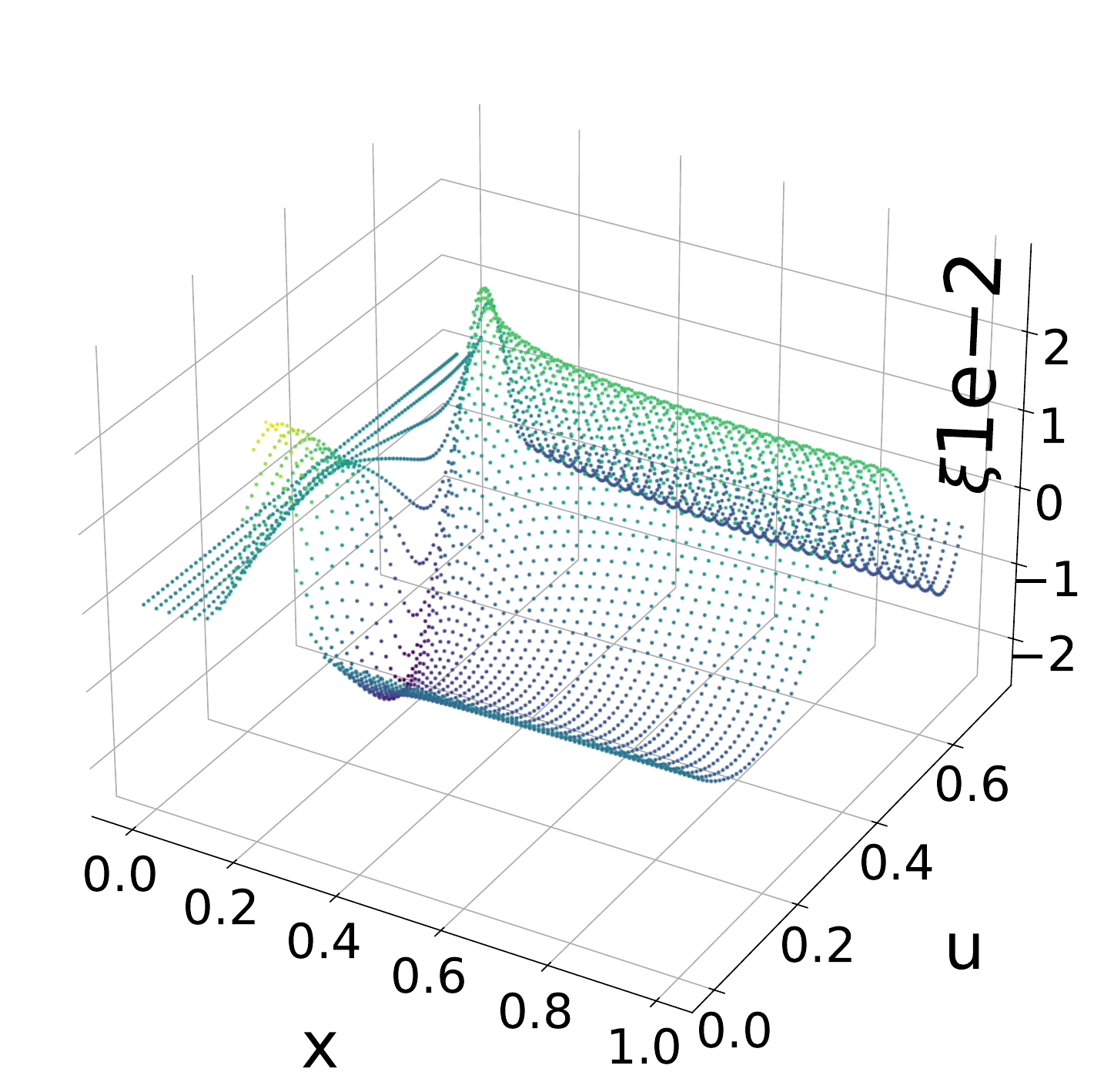}}
	\caption{$\xi=W-1$ throughout the evolution. The top panel shows a subcritical evolution with initial amplitude~$\mathrm{A}=0.001$ and the lower a supercritical evolution with~$\mathrm{A}=0.089$ for $N=400$. For clarity only a subset of the data points are shown.}
	\label{fig:ymsubandsupevols}
\end{figure}

We now turn to analyze different evolution regimes of a Yang-Mills field collapse. As an illustrative example, in Figure~\ref{fig:ymsubandsupevols} we present the field~$\xi(u,x)$ for both subcritical and supercritical initial data. We see that the subcritical evolution disperses completely at the end of the evolution whereas the supercritical evolution stops at earlier times, when our compactness criteria is met. To close in on the threshold of collapse we perform bisection searches, as is common, for grid setups with different resolutions. The results of our bisections are summarized in Table~\ref{table:bisection-params}.

\begin{table}[t!]
\centering
\begin{tabular}{ccccc}
\hline
N    & $\mathrm{A}_*$ & $u_*$ & $\Delta$ \\ \hline \hline

2$\cdot10^3$ & 0.088640998& 1.055$\pm$0.002    &   0.73$\pm$0.02       \\

4$\cdot10^3$ & 0.088640996  & 1.0538$\pm$0.0002     & 0.735$\pm$0.004   \\

6$\cdot10^3$ & 0.088640995 & 1.0537$\pm$ 0.0001     &    0.737$\pm$0.002   \\
\end{tabular}
\caption{Critical parameters obtained for the same family of initial data. All values presented are calculated for grids with different basis resolutions~$N$. $\mathrm{A}_*$ denotes the critical amplitude, $u_*$ denotes the accumulation time and $\Delta$ is the self similarity echoing period defined in Equation~\eqref{eq:deltaapprox}. To calculate $u_*$ and~$\Delta$ we average the possible pairs of zero-crossings (excluding the first echo) and use the standard deviation as the uncertainty. The evolution is fine-tuned up to 9 decimal places.}\label{table:bisection-params}
\end{table}

Regarding the results shown in Table~\ref{table:bisection-params}, we first notice that the critical parameters obtained for each resolution are consistent. Using Equation~\eqref{eq:deltaapprox}, we estimate an echoing period of~$\Delta\approx0.73$. At the continuum level, there's one single critical amplitude value. However, the numerical error depends on the resolution considered, so as we use more gridpoints, the runs become more accurate (in practice, `less dissipative') and we are better able to better resolve runs with stronger initial data.

In~\cite{Gundlach_1997} perturbative arguments are used to calculate the echoing period, and it is found that~$\Delta = 0.73784 \pm 0.00002$, consistent with the value we report in Table~\ref{table:bisection-params}. Additionally, the work of~\cite{Maliborski_2018} estimates a value of~$\Delta = 0.7364 \pm 0.0007$, and~\cite{Choptuik_1996} estimates~$\Delta\approx0.74$. The error estimation here is non-trivial as there is both numerical error and also error from the method used to estimate~$\Delta$, so that it is easy to underestimate the error associated in these estimates. Nevertheless, the results we pointed out are consistent with our findings, despite using very different continuum and numerical setups. 

In the next subsections we discuss local and global self-similarity of critical solutions, then black hole mass scaling and finally present evidence that our results are universal with respect to initial data.

\subsection{DSS Behaviour in Local and Global Quantities}
\label{sec:sshbehaviour}

In this section, we investigate how local and global quantities behave in evolutions that approach the critical solution. Following~\cite{Gundlach_2019}, we observe that~$\chi$ (recalling that $W=1+x^2 \chi \equiv 1+\xi$) is not compatible with exact \ac{DSS}. The critical solution observed should instead be of the form
\begin{align}
    \Tilde{\chi}=e^{-T} \chi,
\end{align} 
where~$T$ is the similarity time.

\begin{figure}[t!]
\centering
\includegraphics[width=\columnwidth]{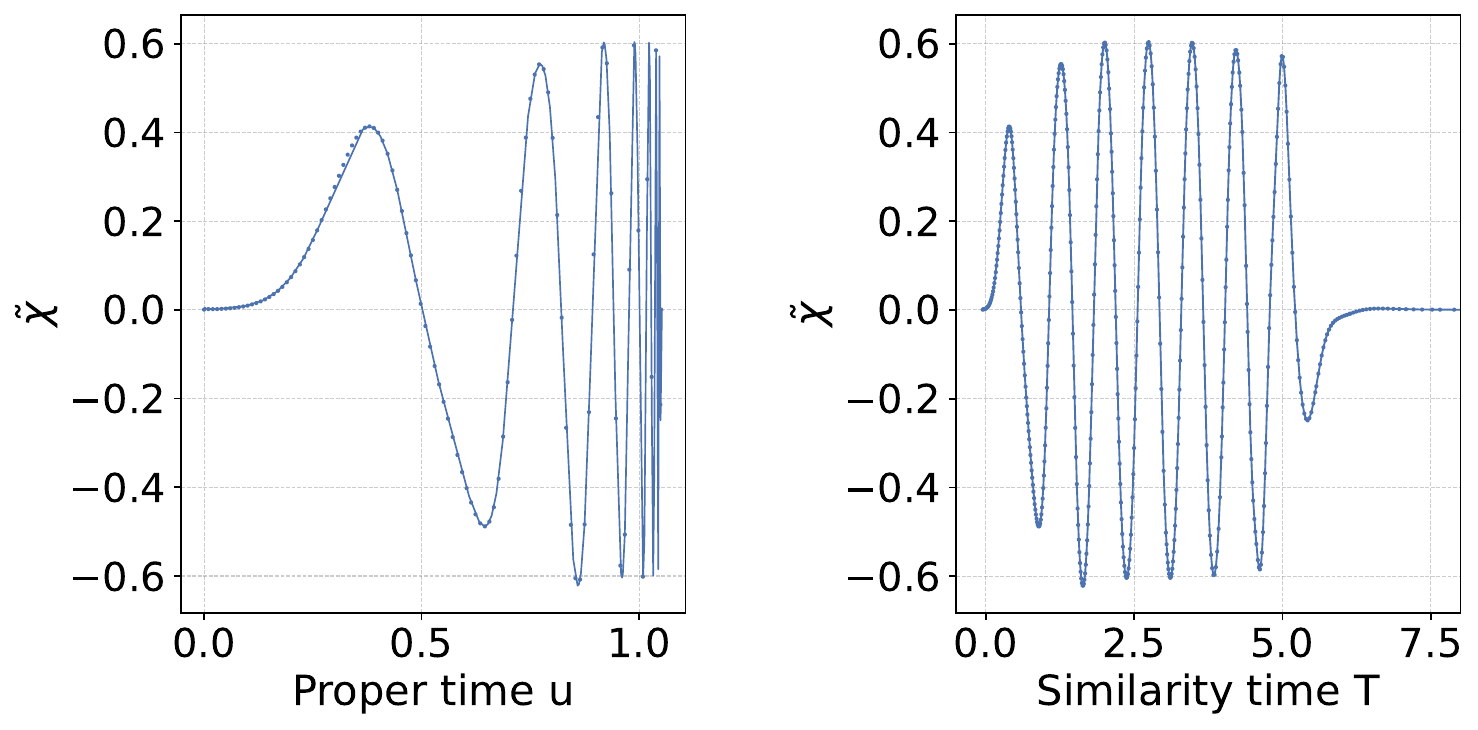}
\caption{$\Tilde{\chi}$ at the origin as a function of proper time~$u$ and similarity time~$T$ for a run of N=$6\cdot10^3$, tuned to~$9$ decimal places, with $\mathrm{A}_*=0.088640995$.}
\label{fig:chitilde}
\end{figure}

We perform a bisection search with~$N=6\cdot10^3$ points and find a critical amplitude of~$\mathrm{A_*} = 0.088640995$, tuned to 9 decimal places. We find that, for this resolution, further tuning our solution did not improve the number of echoes observed. Observe This result is achieved without the use of mesh refinement. In Figure~\ref{fig:chitilde} we show the value of~$\Tilde{\chi}$ at the origin as a function of proper time~$u$ and of similarity time~$T$, for our best tuned run. We see that indeed~$\Tilde{\chi}$ is \ac{DSS} with an echoing period of~$\Delta \approx 0.737$. 

Although we can observe a significant number of echoes for the Yang-Mills field without using mesh refinement, this is much more difficult for the scalar field collapse, as the echoing period is around five times larger in that case.

The shape we find of the solutions near-criticality agrees with the picture found in previous works, that the features of critical solutions in a region around the origin\cite{Maliborski_2018,Choptuik_1996}. Our main goal is, however, to inspect these features at~$\mathscr{I^+}$.

\begin{figure}[t!]
\centering
\includegraphics[width=\columnwidth]{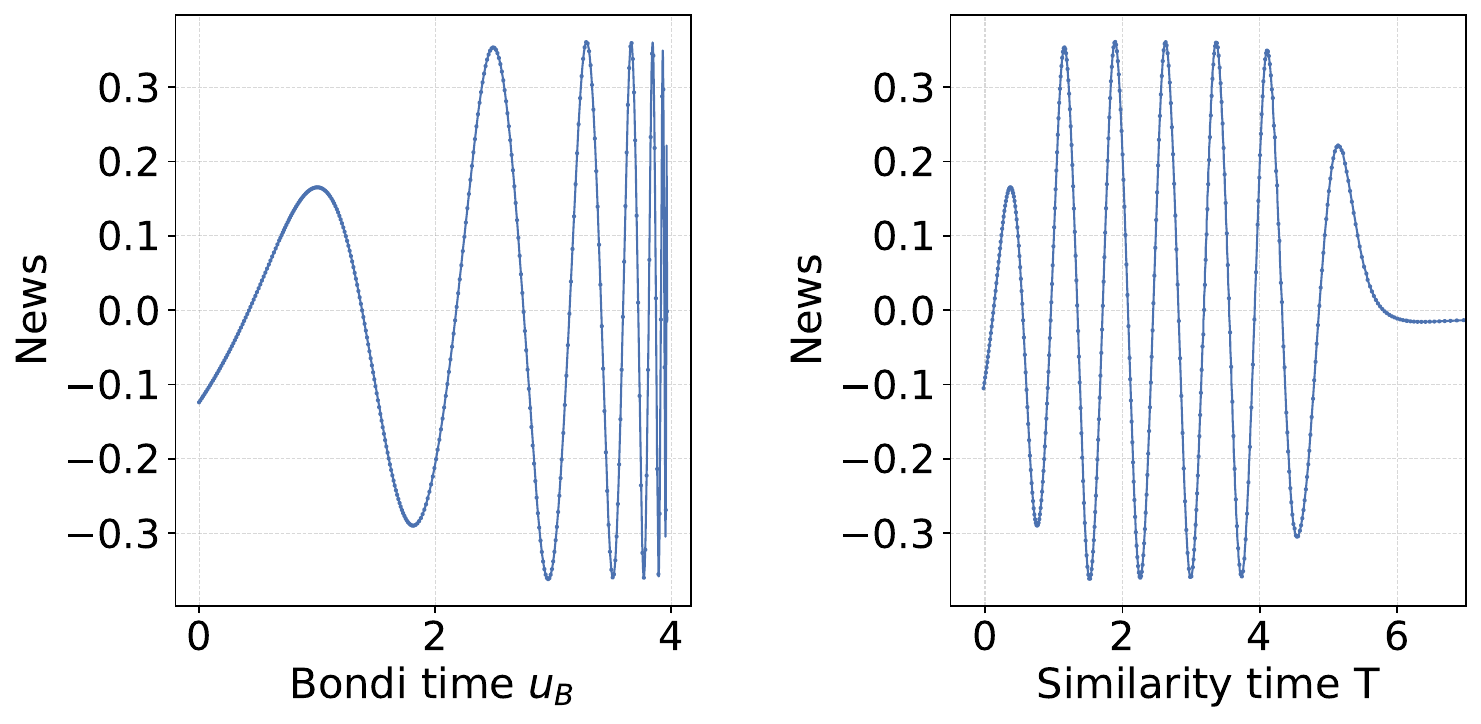}
\caption{News function at $\mathscr{I^+}$ as a function of Bondi time $u$ and similarity time $T$ for a run of N=$6\cdot10^3$, tuned to 9 decimal places, with $\mathrm{A}=0.088640995$.}
\label{fig:news}
\end{figure}

In fact, the \ac{SSH} is the past light cone of the accumulation time. Hence, it is the region of spacetime where we expect that the solution is self-similar. Moreover, since we are using a characteristic foliation of spacetime, we are able to show that the dynamics of the critical solution, and particularly discrete self-similarity, are radiated to~$\mathscr{I^+}$. This is shown in Figure~\ref{fig:news}, where the news function is plotted. An observer at~$\mathscr{I^+}$ is then able to observe features of critical collapse. These are imprinted also in the Bondi mass, the other asymptotic quantity discussed in Section~\ref{asymptotic-quantities}.

\begin{figure}[t!]
\centering
\includegraphics[width=\columnwidth]{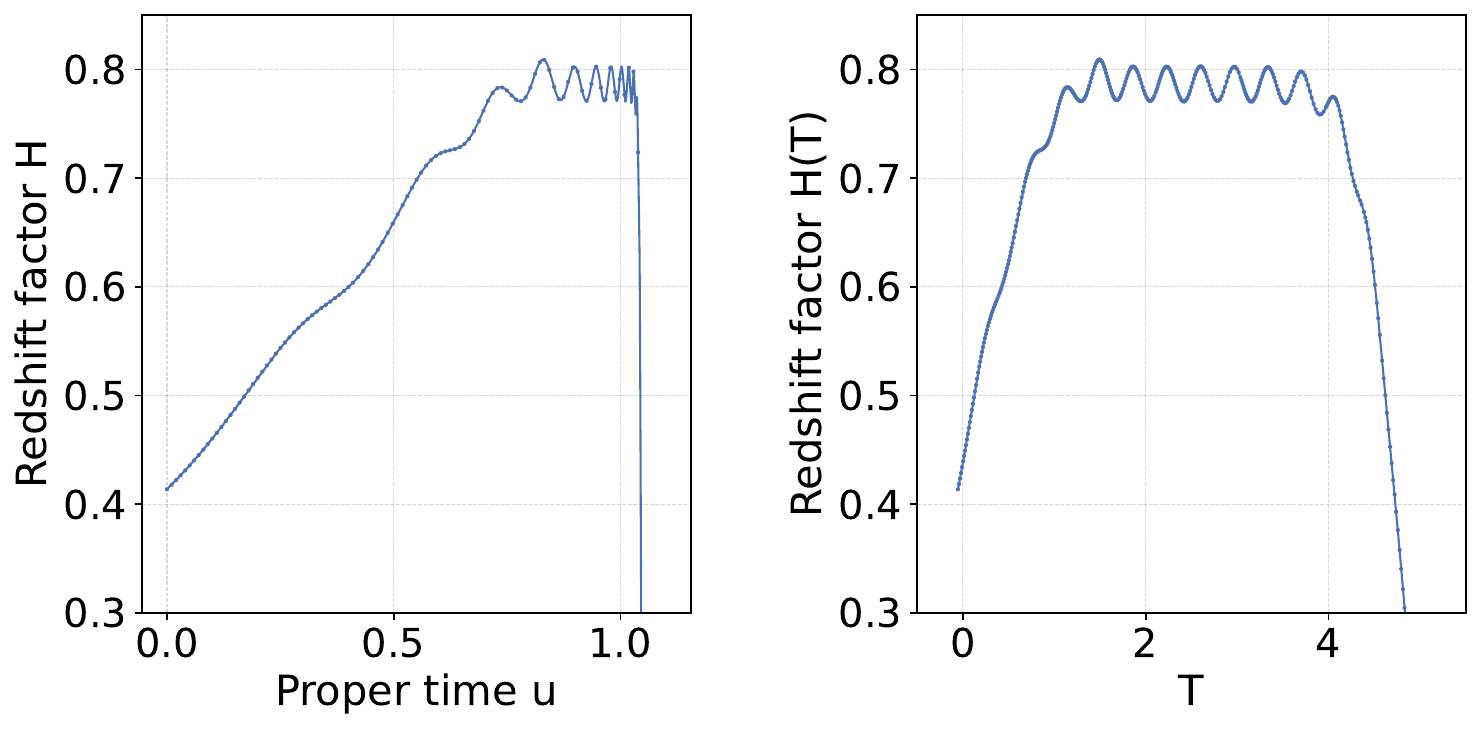}
\caption{Redshift factor~$H(T)=\beta(u_c,\infty$ as a function of Bondi time~$u$ and similarity time~$T$ for a run of~N=$6\cdot10^3$, tuned to 9 decimal places, with~$\mathrm{A}_*=0.088640995$.}
\label{fig:redshift}
\end{figure}

The redshift factor~$H=\beta(u_C,\infty)$ --- introduced in Section~\ref{asymptotic-quantities} --- is shown in Figure~\ref{fig:redshift} for a barely supercritical evolution. This quantity also shows discrete self-similarity as the outgoing light rays that approach~$\mathscr{I^+}$ are themselves subject to an oscillating compactness~$2m/r$, caused by the near-critical collapse of the Yang-Mills field.

\begin{figure}[t!]
\centering
\includegraphics[width=\columnwidth]{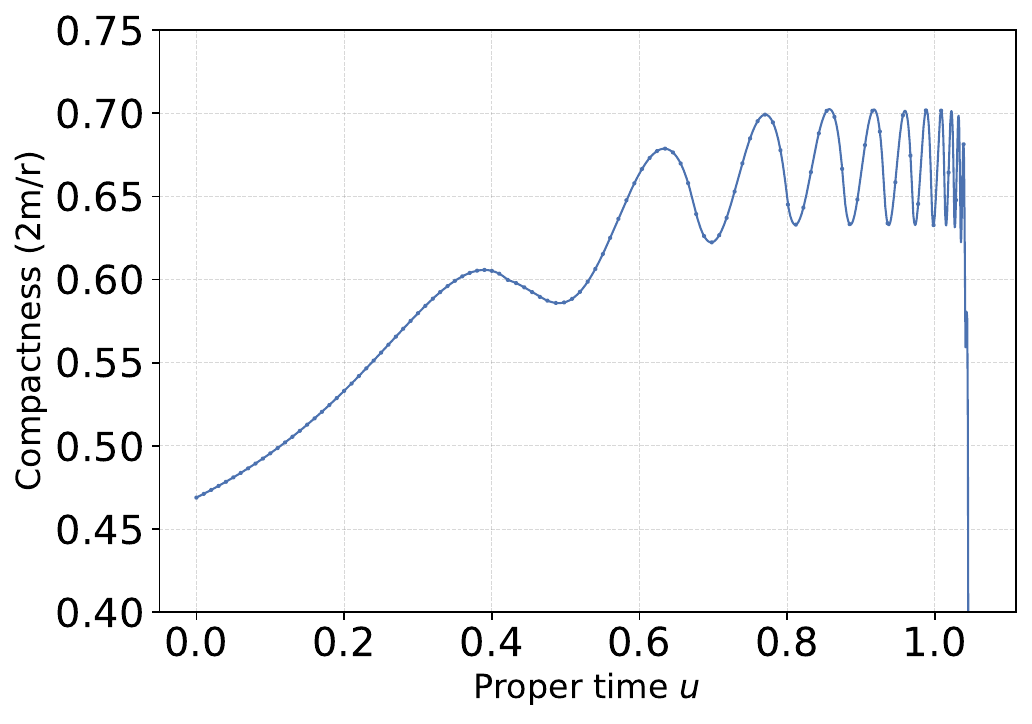}
\caption{Compactness~$\max\left(\frac{2m}{r}\right)$ for a near-critical evolution as a function of similarity time~$T$. The data shown corresponds to a run of~N=$6\cdot10^3$ points, tuned to 9 decimal places, with~$\mathrm{A}=0.088640995$.}
\label{fig:compactness}
\end{figure}

Figure~\ref{fig:compactness} shows that indeed the compactness~$2m/r$ oscillates in a near-critical evolution, caused by the near-critical collapse of the Yang-Mills field.

To summarize, our near-critical solutions share the same universal features as the critical solution, both at the origin and at~$\mathscr{I^+}$. Lastly, our parameter estimates are consistent with those from previous numerical works and with the perturbative results of~\cite{Gundlach_1997}.

\subsection{Black Hole Mass and Curvature Scaling}

As discussed in Section~\ref{intro}, the mass of the black holes formed in evolutions that are barely supercritical has been found to follow a power-law described by Equation~\eqref{eq:massscaling}. For the case of the spherical Yang-Mills, this scaling has been derived by perturbing the critical solution~\cite{Gundlach_1997}, yielding a critical exponent of~$\gamma = 0.1964 \pm 0.0007$.

\begin{figure}[t!]
\centering
\includegraphics[width=\columnwidth]{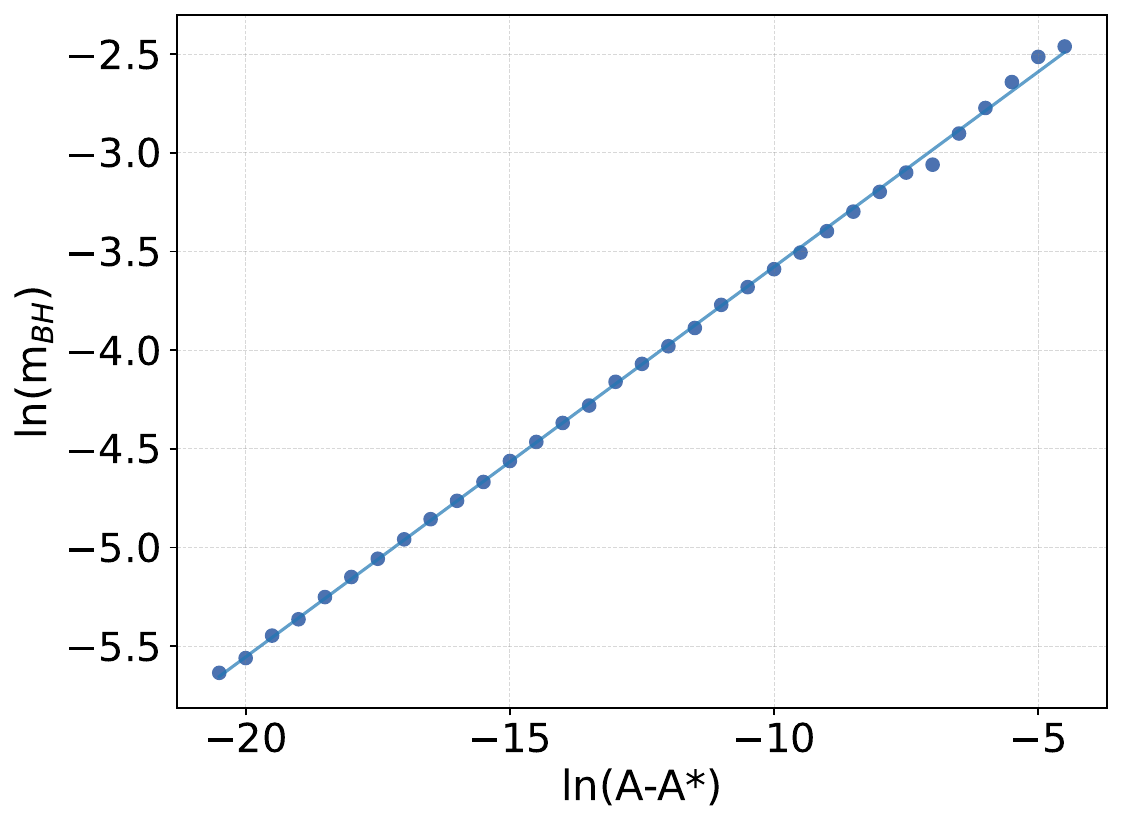}
    \caption{Scaling of the final black hole mass as a function of the distance to the critical amplitude $\mathrm{A}_*$ for supercritical evolutions. We fit the data and find a scaling of the black hole mass~$\ln(m_{\mathrm{BH}})\simeq0.1977\ln(\mathrm{A}-\mathrm{A_*})-1.6$, yielding a critical exponent of $\gamma=0.1977\pm0.0009$. Each point in this figure corresponds to the final black hole mass of an evolution with~$4\cdot10^3$ gridpoints.}
\label{fig:massscaling}
\end{figure}

In this Section, we restrict ourselves to the critical solution previously obtained for~$4\cdot10^3$ gridpoints, with~$\mathrm{A}_*=0.088640996$, as shown in Table~\ref{table:bisection-params}. In Figure~\ref{fig:massscaling}, we plot the values of the estimate mass of the black hole, as a function of the distance to the critical parameter.

\begin{figure}[t!]
\centering
\includegraphics[width=\columnwidth]{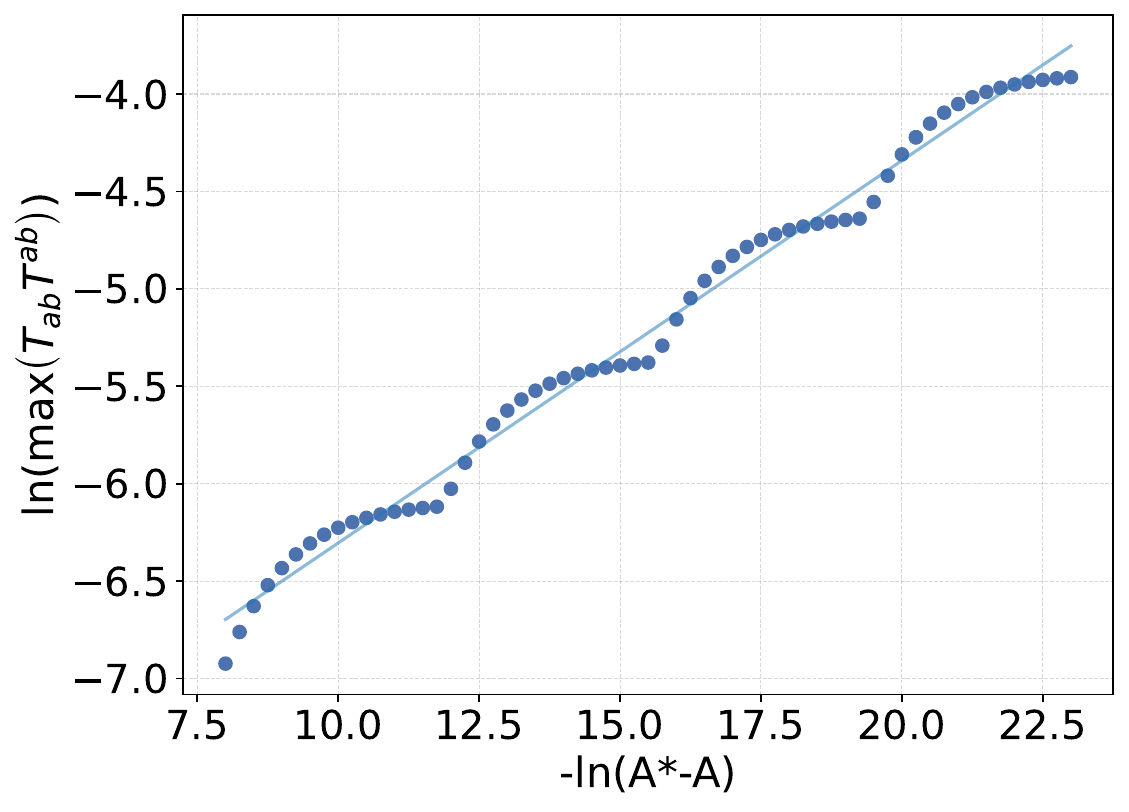}
    \caption{Scaling of~$\max(\mathrm{T_{ab}}\mathrm{T^{ab}})$ as a function of the distance to the critical amplitude~$\mathrm{A}_*$ for subcritical evolutions. We fit the data and find a scaling law of~$\ln(\max(\mathrm{T_{ab}}\mathrm{T^{ab}}))\simeq0.197\ln(\mathrm{A}-\mathrm{A_*})-8.263$, yielding a critical exponent of~$\gamma=0.197\pm0.002$. Each point in this figure corresponds to~$\max(\mathrm{T_{ab}}\mathrm{T^{ab}})$ from an evolution with~$4\cdot10^3$ gridpoints.}
\label{fig:subcriticalscaling}
\end{figure}

\begin{figure}[t!]
\centering
\includegraphics[width=\columnwidth]{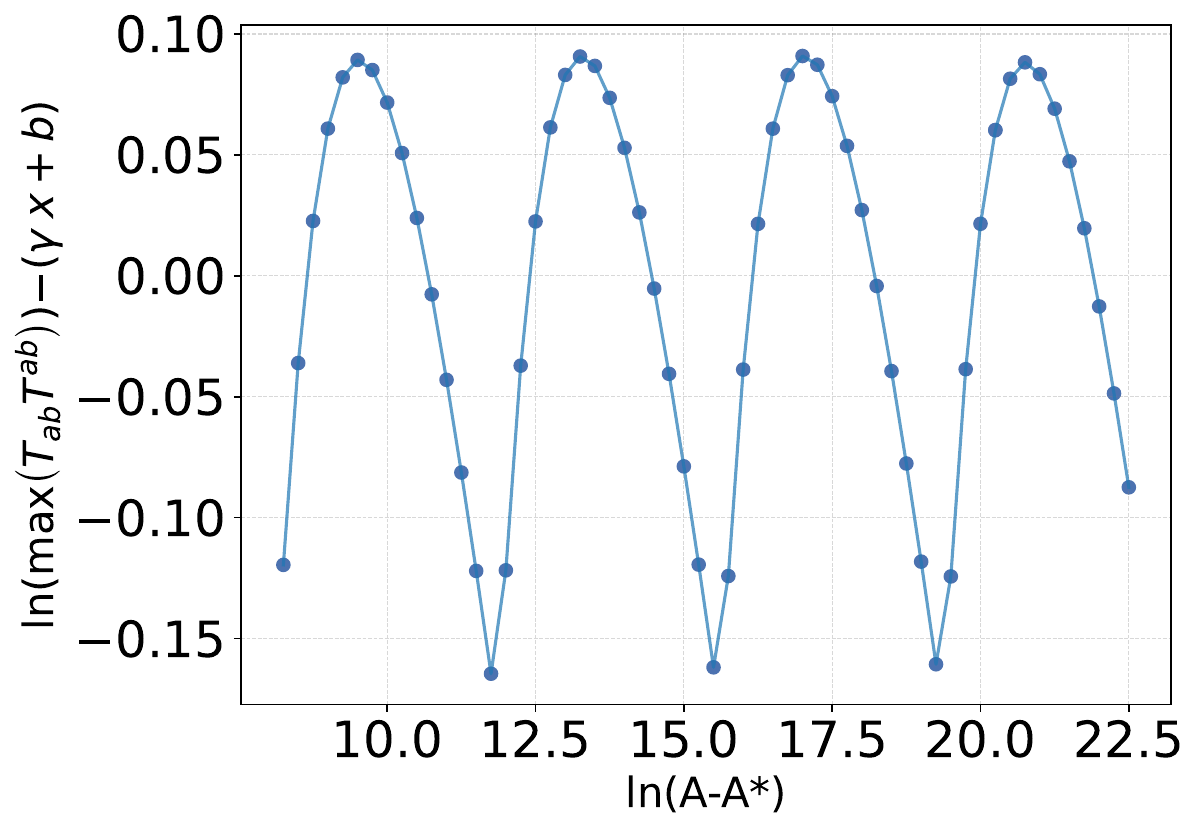}
    \caption{Here we plot the same data from Figure~\ref{fig:subcriticalscaling}, after subtracting the linear fit~$0.197\ln(\mathrm{A}-\mathrm{A_*})-8.263$. The underlying periodic behavior in the scaling of~$\max(\mathrm{T_{ab}}\mathrm{T^{ab}})$ is very evident.}
\label{fig:scalingminusfit}
\end{figure}

We fit the data present in Figure~\ref{fig:massscaling} with an ansatz~$\ln(m_{\mathrm{BH}})=\gamma\ln(\mathrm{A}-\mathrm{A_*})+\mathrm{b}$ (see Equation~\eqref{eq:massscaling}) and find that the final black hole mass scales as~$\ln(m_{\mathrm{BH}})=0.1977\ln(\mathrm{A}-\mathrm{A_*})-1.6171$. We then extract a critical exponent of~$\gamma=0.1977$. This value is again compatible with that from perturbation theory in~\cite{Gundlach_1997}, but and also with the values estimated in earlier numerical studies. In particular~\cite{Maliborski_2018} finds an exponent~$\gamma = 0.19714 \pm 0.00074$, and~\cite{Choptuik_1996} estimates~$\gamma = 0.20$, both using Cauchy formulations.

This same scaling behavior happens for barely subcritical evolutions. In this scenario, we compute the maximum value of~$\mathrm{T_{ab}}\mathrm{T^{ab}}$, which is a proxy for the maximum curvature within an evolution. In Figure~\ref{fig:subcriticalscaling}, we plot this quantity as a function of the distance to~$\mathrm{A_*}$, and extract a scaling law of~$\ln(m_{\mathrm{BH}})\simeq0.197\ln(\mathrm{A}-\mathrm{A_*})-8.263$, with~$\gamma=0.197\pm0.002$. This exponent is consistent within errors with the~$\gamma$ extracted from supercritical evolutions, as well as with those from~\cite{Choptuik_1996,Gundlach_1997} and~\cite{Maliborski_2018}. In Figure~\ref{fig:scalingminusfit}, we plot the same data from Figure~\ref{fig:subcriticalscaling}, but with the linear fit subtracted, revealing the underlying periodic behavior.

We recall that the values of the black hole masses are estimated by taking the value of the Misner-Sharp mass at the gridpoint in which the compactness reaches our criterion to mark an evolution as supercritical, without interpolation between gridpoints. In fact, assuming the Cosmic Censorship Conjecture holds true, there will still be mass-energy left outside of the compactness peak that will later fall through the horizon, and thus this estimate will indeed not correspond to the true final black hole mass. This, together we the fact that we do not interpolate to estimate~$\mathrm{m_{BH}}$, could wash out fine structure, which may explain why we do not see the periodic behavior on the supercritical side in Figure~\ref{fig:massscaling}. Nevertheless, this approach is sufficient to observe the scaling as predicted with perturbation theory, with a critical exponent~$\gamma=0.1977$, in agreement with theoretical computations and numerical estimations from previous works.
If we take the closest simulation to the critical solution on the supercritical side, we can get an estimate of the smallest black hole we can build for this family of initial data, at this resolution. We find that the smallest black hole is~$2.75\%$ of the initial Bondi mass in that evolution. At the time the horizon forms, about~$71\%$ of the Bondi mass is contained within the Black Hole.

\subsection{Bondi Mass Decay}
\label{bondimassdecayym}

\begin{figure}[t!]
\centering
\includegraphics[width=\columnwidth]{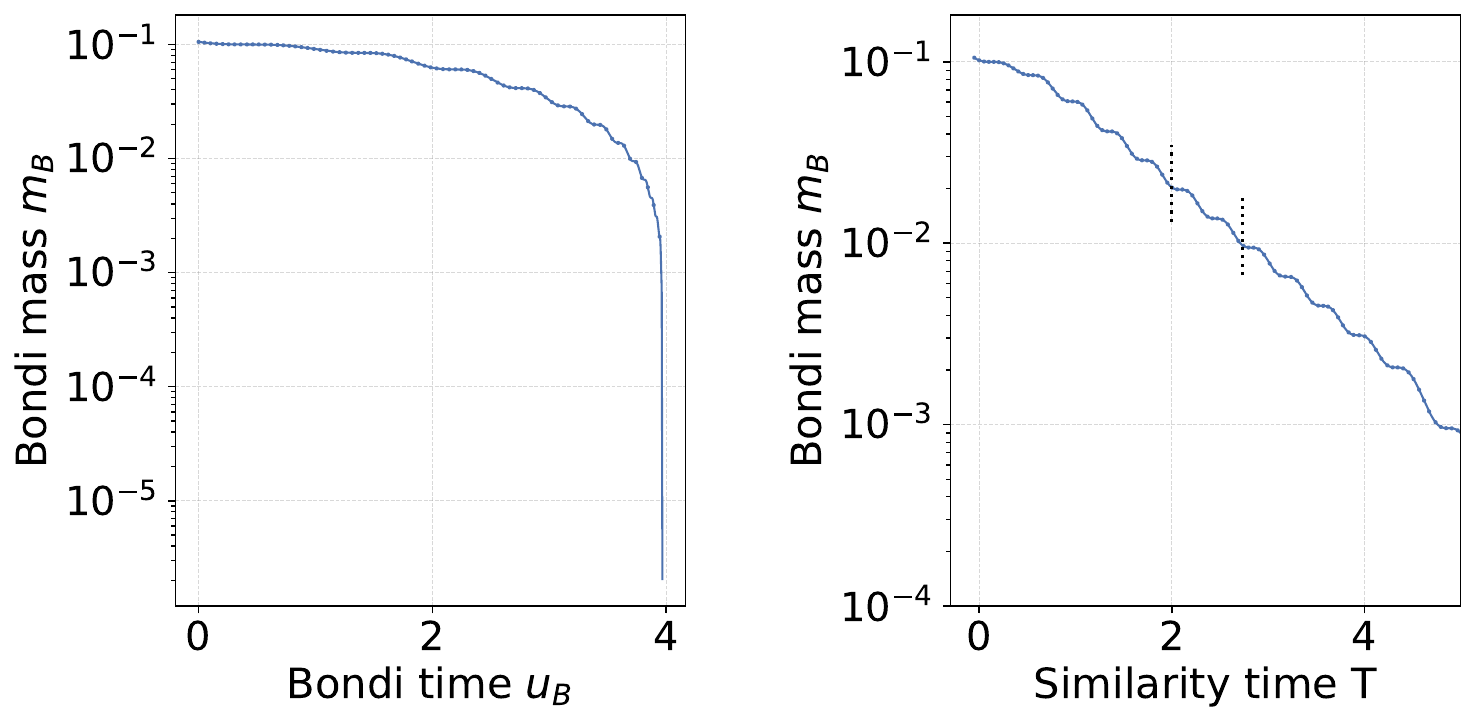}
\caption{Bondi mass as a function of Bondi time $u_B$ and similarity time $T$. The data here presented corresponds to the same run as Figure~\ref{fig:chitilde}. We find that the Bondi mass in the last time step of this evolution ($\mathrm{T_F}$) is $\mathrm{m_{B,T_F}}=1.9*10^{-5} \cdot \mathrm{m_B(t=0)}$. The right panel shows a zoom-in of the region where the periodicity is visible. The vertical lines are spaced in 0.73, which is approximately the same value we calculated for~$\Delta$.}
\label{fig:bondimassdecay}
\end{figure}

As discussed in Section~\ref{asymptotic-quantities}, the Bondi mass is an asymptotic quantity which measure the total mass present on an outgoing null hypersurface. We also observe this global quantity to be \ac{DSS} (see Figure~\ref{fig:bondimassdecay}), and thus features of the critical collapse which happen around the center of spherical symmetry are radiated towards future null infinity. In fact, the period at which the Bondi mass is shown to decay is approximately the same as the one which is estimated for the echoing around the region of the center of the collapse.

\begin{figure}[t!]
\centering
\includegraphics[width=0.8\columnwidth]{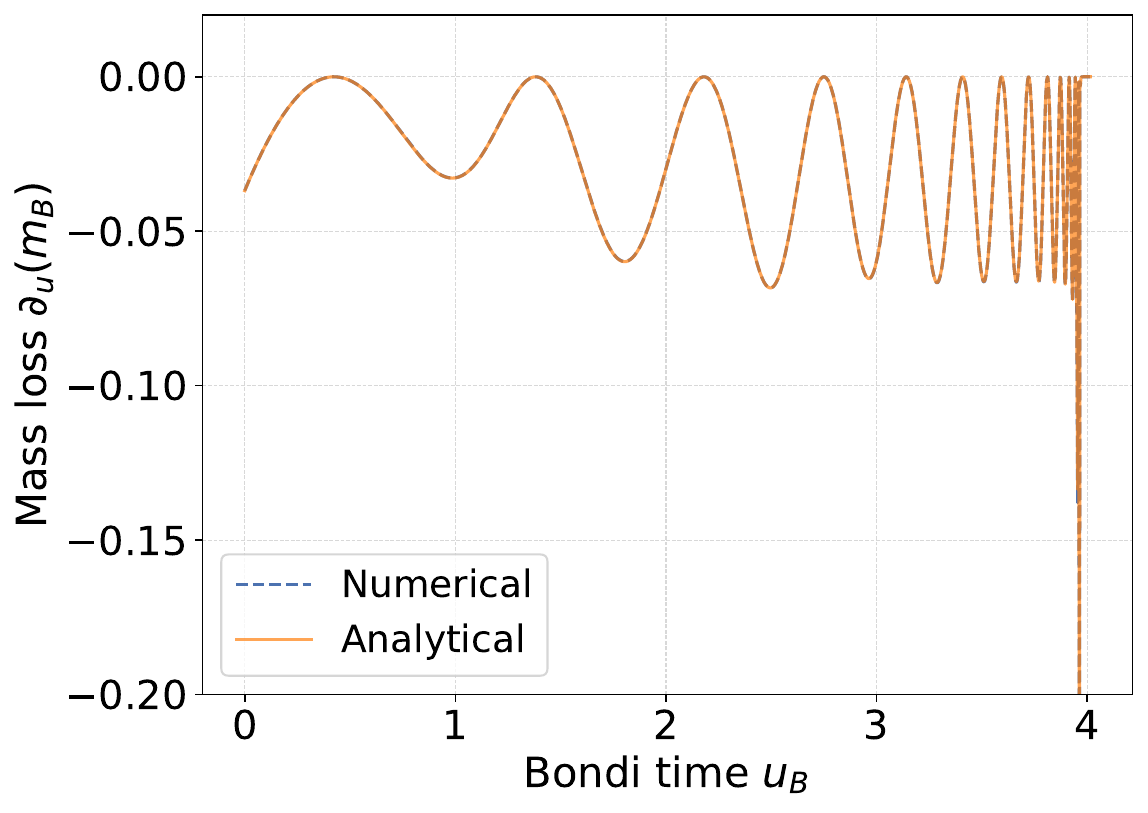}
\caption{We compare the mass loss at $\mathscr{I^+}$ for the Yang-Mills collapse code. The numerical values refer to the values obtained by taking a time derivative of $m$ solved throughout our evolution. The analytical value refers to the values computed using a derived equation for the mass loss at $\mathscr{I^+}$. The data here presented corresponds to the same run as Figure~\ref{fig:chitilde}.}
\label{fig:masslossym}
\end{figure}

The mass loss $\partial_u(m_{\mathrm{B}})$ at $\mathscr{I^+}$ throughout the evolution is expressed by Equation~\eqref{eq:masslossanalyticalym}, which was obtained from the $\{r,r\}$ Einstein equation, combined with our $m$ and $\beta$ hypersurface equations. In Figure ~\ref{fig:masslossym}, we plot the numerical results obtained by taking a time derivative of the variable~$m$, computed with our numerical scheme, along the value we obtain analytically from the Einstein equations. We see that these values correspond almost exactly to the analytical expected result.

This is in fact a non-trivial test of our numerical scheme since we don’t directly solve all Einstein equations, and in particular we don’t solve equation~\eqref{eq:masslossanalyticalym}. Instead, we calculate~$m$ in slices of constant~$u$ by integrating out an expression that depends on the remaining evolution variables. This provides additional evidence that our results and implementation are valid.

\subsection{Universality}

So far, we have been studying families of initial data that are described by Equation~\eqref{eq:initial-dataym}, in which we fix~$r_0=0.3$ and~$\sigma=0.1$. We have studied the influence of changing the amplitude~$\mathrm{A}$ and fine-tuning to its critical value~$\mathrm{A}_*$. We now study the collapse of different families of initial data, in particular by considering initial data with different~$\sigma$ and~$r_0$ values. For several combinations of~$\{\sigma,r_0$\} values, we performed a bisection search to find the critical amplitude, and the associated power-law and scaling period parameters. These are shown in Table~\ref{table:universality-params}.

As expected, runs with initial data that is more localized (with smaller~$\sigma$) have a smaller accumulation time, meaning that the apparent horizon starts to form at an earlier time in the evolution. Moreover, the critical amplitude is smaller in the evolutions with denser initial data. We obtain the same  picture as in the Section~\ref{sec:sshbehaviour}, with threshold solutions characterized by the same echoing period~$\Delta$, despite using different initial data. This gives supporting evidence that, regardless of the initial data profile, a near-critical solution will `lose' its features along the evolution and approach the critical solution.

\begin{table}[t!]
\centering
\begin{tabular}{ccccc}
\hline
$r_0$ & $\sigma$   & $\mathrm{A}_*$ & $u_*$ & $\Delta$ \\ \hline \hline

0.3 & 0.1  & 0.08864100    & 1.05375286       & 0.73746089   \\
0.3 & 0.08  & 0.06977534    & 0.95649628        & 0.70009564   \\
0.4 & 0.1  & 0.06512129    & 1.23750094        & 0.73611316   \\
0.4 & 0.08  & 0.05111100    & 1.14037085        & 0.73883653   \\

\end{tabular}
\caption{Critical parameters obtained for the same family of initial data, with different grid setups and code accuracy. All values presented are calculated for grids with resolution of~$N=2\cdot10^3$. $\mathrm{A}_*$ denotes the critical amplitude, $u_*$ denotes the accumulation time and $\Delta$ is the self similarity echoing period defined in Equation~\eqref{eq:deltaapprox}. To calculate $u_*$ and $\Delta$ we use the 2nd and 3rd echos. All evolutions are fine-tuned up to 8 decimal places.}
\label{table:universality-params}
\end{table}

\section{Conclusions}
\label{sec:conclusions}

In this work, we examined the critical collapse of Yang-Mills field using a null foliation of spacetime in spherical symmetry with a purely magnetic ansatz. Our main goals were to use a formulation that would allow the study of this process from the point of view of an observer located at future null infinity and  to achieve fourth-order accuracy. This allowed for the computation of radiation quantities, and was achieved by virtue of compactified Bondi coordinates.

The evolutions that are near-critical exhibit features of the critical solution of the Yang-Mills field. They are found to be \ac{DSS} with an echoing period of~$\Delta\approx0.7388$. In barely supercritical evolutions, the mass of the black holes formed was shown to follow a power-law in relation to the distance to the critical parameter. This relation is characterized by a critical exponent of~$\gamma=0.1977$ in agreement with earlier work. We also presented evidence that near the threshold of collapse, all solutions approach the critical solution, which is thus universal.

The critical parameters estimated are not only consistent with the computations of \cite{Gundlach_1997}, but also agree with previous numerical works that were restricted to the strong field region. Our key finding is that, as in the case of scalar field collapse, known features of the critical solution in the strong-field region, such as the echoing period~$\Delta$, are inherited at future null infinity.

Possible future directions consist in exploring dropping the magnetic ansatz to investigate all the degrees of freedom of the Yang-Mills connection in spherical symmetry, as explored in~\cite{Maliborski_2018} with null infinity. Another promising direction would be to drop spherical symmetry, as has been done for scalar fields in~\cite{Choptuik:2003ac,Choptuik:2004ha,Baumgarte_2018,marouda2024,Gundlach_2024}. The study of aspherical perturbations of spherical critical solutions from a matter model with a small echoing period is desirable, as it is easier to study perturbations given many periods of periodic data. This will most likely yield more accurate results than using a scalar field, as one needs much less resolution to observe the same critical phenomena. Recent work~\cite{Gundlach:2024mld} presents a roadmap for critical collapse simulations in null coordinates of nonspherical data, and the treatment by Cauchy evolution could be performed with more common methods.

Moreover, a limitation of our work consists in the fact that we use compactified Bondi coordinates which become singular at the event horizon. In this way, we are limited to the spacetime exterior to black hole formation. Since we focus on the spherically symmetric case, the horizon forms simultaneously in all radial null directions from the center of collapse. However, in the collapse of nonspherical initial data, the horizon forms at different retarded times for different angles. Thus, for data that is not spherically symmetric, it could be helpful to use a formulation that can penetrate horizons. In~\cite{Crespo_2019}, for example, an evolution algorithm is presented for the \ac{CIVP} based upon an affine parameter rather than the areal radial coordinate, which is potentially applicable to the entire exterior spacetime extending  additionally to the interior of the black hole. The use of carefully chosen coordinates may not only allow for the study of the collapse of non-spherically symmetric data, but also help advance the study of the scaling quantities within barely supercritical simulations.

In closing, our results show that the standard picture of critical collapse for the Yang-Mills field in spherical symmetry is indeed observable at future null infinity, and that the features of the critical solution are preserved in this region of spacetime.

\acknowledgments

We are grateful to Thanasis Giannakopoulos, Sascha Husa and Alex Vañó-Viñuales for helpful conversations and/or feedback on the manuscript. The work is based on the Master thesis of the first Author~\cite{rita2023}. We acknowledge financial support provided by FCT/Portugal through grants UID/99/2025, 2024.03599.BD, and PeX-FCT (Portugal) program
2023.12549.PEX. The authors thankfully acknowledge the computer resources, technical expertise and assistance provided by CENTRA/IST. Computations were performed at the cluster ``Baltasar-Sete-Sóis''  supported by the 2020 ERC Advanced Grant “Black holes:
gravitational engines of discovery” grant agreement
no. Gravitas–101052587 and ERC-Portugal program Project ``GravNewFields''. This study was as well supported by Ministerio de Ciencia, Innovación y Universidades, with funding from European Union NextGenerationEU (PRTR-C17.I1) and the Comunitat Autònoma de les Illes Balears through the Conselleria d'Educació i Universitats  (SINCO2022/6719).
\newpage
\bibliography{main.bbl}
\end{document}